\documentclass[sigconf, nonacm]{acmart}
\usepackage{algorithm}
\usepackage{algorithmic}
\usepackage{svg}
\usepackage{subcaption}

\newcommand\vldbpagestyle{plain}

\begin{document}
\title{Enhancing Data Provenance and Model Transparency in Federated Learning Systems---A Database Approach}

\author{Michael Gu, Ramasoumya Naraparaju, Dongfang Zhao}
\affiliation{%
  \institution{University of Washington}
}
\email{{michgu,soumya22,dzhao}@uw.edu}

\begin{abstract}
Federated Learning (FL) presents a promising paradigm for training machine learning models across decentralized edge devices while preserving data privacy. Ensuring the integrity and traceability of data across these distributed environments, however, remains a critical challenge. The ability to create transparent artificial intelligence, such as detailing the training process of a machine learning model, has become an increasingly prominent concern due to the large number of sensitive (hyper)parameters it utilizes; thus, it is imperative to strike a reasonable balance between openness and the need to protect sensitive information.

In this paper, we propose one of the first approaches to enhance data provenance and model transparency in federated learning systems with practical communication overhead. Our methodology leverages a combination of cryptographic techniques and efficient model management to track the transformation of data throughout the FL process, and seeks to increase the reproducibility and trustworthiness of a trained FL model. We demonstrate the effectiveness of our approach through experimental evaluations on diverse FL scenarios, showcasing its ability to tackle accountability and explainability across the board. 

Our findings show that our system can greatly enhance data transparency in various FL environments by storing chained cryptographic hashes and client model snapshots in our proposed design for data decoupled FL. This is made possible by also employing multiple optimization techniques which enables comprehensive data provenance without imposing substantial computational loads. Extensive experimental results suggest that integrating a database subsystem into federated learning systems can improve data provenance in an efficient manner, encouraging secure FL adoption in privacy-sensitive applications and paving the way for future advancements in FL transparency and security features.

\end{abstract}

\maketitle
\pagestyle{\vldbpagestyle}
\section{Introduction}
\subsection{Background}
Since its widespread adoption in all sectors of technology over the past couple of decades, there has been much interest in developing a data-secure machine learning (ML) system. Federated Learning (FL) ~\cite{zli, yejia, liu1, wei, konecny, verticalFL1} is a revolutionary approach to collaborative ML which enables models to be trained across numerous devices while keeping the data localized. This approach, first introduced in 2016 by Google, was proposed to address the growing concerns of data privacy in traditional ML models, which typically required data to be centralized for processing ~\cite{FedAvg01}. The unique design of FL allowed machine learning models to learn from a vast amount of decentralized data without moving the data from its original source; this addressed the core issues of ML by directly cutting off the aspect of transferring highly sensitive data between a device and a server ~\cite{challenges1}.

Since its inception, FL has worked towards the development of increasingly optimized algorithms for decentralized learning as well as increasing its adoption in various artificial intelligence applications ~\cite{tracer, dai, tan1}. At the core of federated learning research is the study of secure aggregation algorithms for decentralized learning; these algorithms have enabled the decentralized training of deep networks across multiple devices, and are directly responsible for the accuracy and privacy of the training process. These works directly extend towards building truly secure ML, providing valuable insights into maintaining data privacy while enabling effective machine learning in a federated setting ~\cite{SecAGG}. In addition, existing FL frameworks such as TensorFlow Federated ~\cite{TFF}, PySyft ~\cite{PySyft}, and OpenFL ~\cite{OpenFL}, have made similar strides in simplifying the implementation of FL algorithms and efficient communications. These advancements have allowed for further work in optimizing the federated learning process, improving in performance in areas such as accuracy and efficiency ~\cite{FedAvg01, chen}. Not only have they propelled the field of FL forward, but they also have facilitated its successful application in various industrial areas, notably in healthcare and telecommunications ~\cite{medical1, medical2}.

However, due to the nature of FL---where participating parties do not reveal their private training data or models---there has been increasing concern regarding the data provenance and model transparency in FL.
This gap between FL's versatility and its lack of traceability highlights the pressing need for a comprehensive approach that not only tackles the core security issues of FL but also preserves the efficiency and effectiveness of the overall system. This work aims to bolster the ability to audit and verify the training process of FL, demonstrate the feasibility of implementing such a feature, and showcase its minimal if not nonexistent impact on resource overhead, training accuracy, and other relevant ML metrics. 

\subsection{Motivation}
Despite the significant advancements in Federated Learning (FL), several challenges persist, specifically in the direction of ensuring traceability of data and clarity of model operations ~\cite{yang1}. Unsurprisingly, these issues are not trivial; they form the bedrock of trust and reliability in FL systems. Without a clear understanding of how data is used and how models operate, it becomes difficult to fully trust the outcomes of these systems, which is a baseline requirement for most if not all applications of FL.

Several attacks targeting FL systems have been identified since its inception, which invalidated the original ideology that FL systems inherently solved the data privacy issue ~\cite{yin1, geiping1}. One, for example, employs a model inversion attack on generative adversarial networks which showed that federated clients' real-time training data cannot be fundamentally secured. This attack specifically demonstrated that it was possible to reconstruct training datasets from output parameters ~\cite{hitaj}. Another proposes a methodology for client model poisoning, which invalidates the accuracy of a model and effectively backdoors the FL system ~\cite{bagdasaryan}. A third common attack exploits ``the privacy vulnerabilities of the stochastic gradient descent algorithm'', a popular machine learning optimization method used to train neural networks used by systems such as FL ~\cite{shokri}. These attacks, studied extensively by various research groups, underscore the need for enhanced data provenance and model transparency in FL systems. Not only do these attacks compromise the integrity of FL systems, but they also pose significant privacy risks. They highlight the vulnerabilities in current FL systems and the urgent need for more robust security measures. This work seeks to mitigate the risk from such attacks by auditing all data to determine whether or not the existing training is reliable and factual. 

In response to these challenges, various defensive mechanisms have been proposed, such as differential privacy (DP) ~\cite{diffprivateadmm, Abadi} and secure multi-party computation (MPC) ~\cite{mpc1, mpc2, mpc3, mpc4}. While these defenses provide some level of protection, they exhibit various limitations for practical applications. The trade-offs they introduce can inhibit the effectiveness and practicality of FL systems, making it a challenging task to balance security with performance. For example, the communication cost of MPC increases with the number of parties involved and relies on all parties to act honestly, similar to the issues faced by federated learning itself. On the other hand, DP introduces noise to the data, which degrades the quality of training results. 

\begin{figure}[t]
    \centering      
    \includegraphics[width=\linewidth]{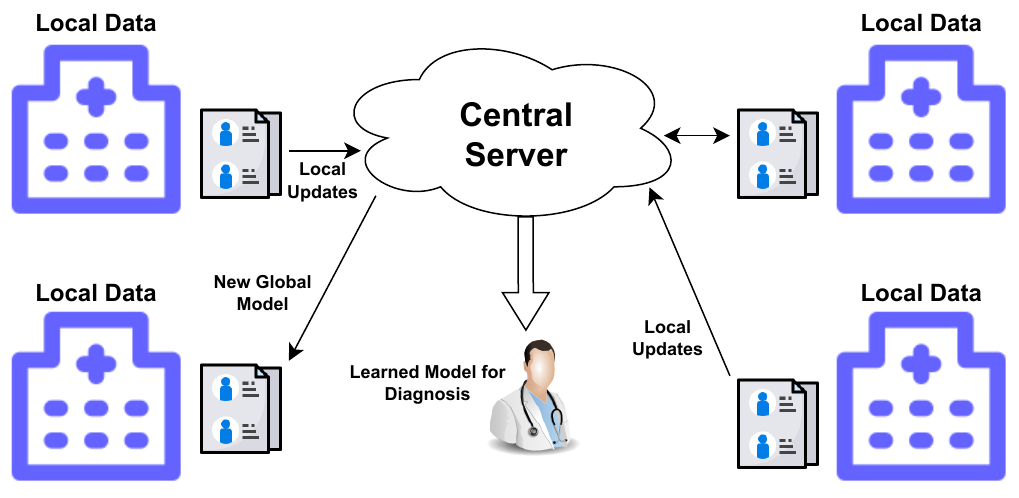}
    \caption{Application of federated learning in healthcare: Diagnosing disease via learning from medical records distributed from regional hospitals}
\end{figure}

The lack of auditability in FL systems has also been a major point of criticism, as it is crucial for ensuring accountability during and after training. Without it, it becomes extremely challenging to verify whether the data and the learning process adhere to required data regulations, and whether or not the training results prove trustworthy. This lack of auditability, coupled with the challenges presented earlier, hinder the wider adoption and acceptance of FL systems. It raises questions about efficiency, accuracy, and compliance with privacy requirements, which goes against the original purpose it was designed for. 

Our approach to tackling these issues is fundamentally different; we aim to enhance data provenance and model transparency by providing a system that audits the data and highlights potential issues during training. Instead of focusing on solely preventing these problems, where there is much work already done, we work towards empowering users to have the ability to understand and scrutinize data and model behavior, fostering accountability and transparency in FL. This work addresses these gaps and comprehensively enhances the data provenance and model transparency in FL systems. Not only do we work towards improving the technical aspects of FL, we also aim to build public trust in these systems, ensure its ethical use, and fully realize its potential as a novel and revolutionary collaborative machine learning paradigm.

\subsection{Proposed Approach}
Federated Learning (FL) systems are often seen as black-box models due to the inherent nature of machine learning and the usage of privacy-preserving methods. This lack of transparency leads to difficulty in assessing model fairness and interpreting model behavior. We propose a comprehensive methodology that allows for improved auditing of machine learning models using a data-decoupled FL architecture that stores all relevant data from the training process, including snapshots of each client's local model.

In addition to enhancing transparency, we also propose a novel approach to verifying the training of FL clients by using chained cryptographic hashing to both track and verify the integrity of local models at each step of training. Each piece of data involved in the training process is hashed and linked in a chain, creating a verifiable and tamper-proof record of data usage reminiscent of blockchains. With this technique, we not only ensure data integrity but also allow for the creation of trustworthy FL models, as reproducing a verified edge device's training yields an exact match of the stored hash.

To our knowledge, this approach is one of the first attempts to bridge the gap between artificial intelligence transparency and FL black-box training. By ensuring model transparency and data provenance, we make FL systems more understandable, verifiable, and trustworthy. The unique usage of cryptographic techniques for data provenance opens up further application of cross-field principles to FL/ML research and supports greater security in such applications.

\subsection{Contributions}
 Our work spans across various facets of FL, including training verification, data provenance, and system architectures, which have collectively improved the transparency and reliability of the training process. The following sections provide a detailed understanding of our work and its implications for the field of federated learning.

\subsubsection{Model Provenance in Databases}
Our research introduces the implementation of a data provenance feature, an additional defensive layer on top of the traditional line of FL security research which proposes hard and fast preventative measures. This feature involves the strategic storage of model snapshots in databases at each round of the training process. Each model snapshot serves as a comprehensive evolution of each client model over time, providing a clear lineage of how the model has learned from the aggregated data of all participating clients further enhancing the transparency of an FL system. In turn, the improved transparency facilitates a more detailed analysis and understanding of the learning process, enabling auditors and associated parties to gain insights into the model's development and performance at various iterations of the process. 

In addition to the storage of model snapshots, the use of persistent databases to store and manage these snapshots brings about several practical benefits. Primarily, it ensures efficient and reliable access to the model's history. Furthermore, it facilitates the management of large volumes of data, a common challenge in large-scale FL applications ~\cite{polyzotis}. By addressing these issues, our data provenance feature can significantly improve the traceability and transparency of FL systems, building confidence in the results of any and all data audits performed.

\subsubsection{Chained Hashing for Training Verifiability}
While we tackle data transparency with the introduction of model snapshot storage, we also aim to tackle FL authenticity via chained hashing methods. Our proposed method uses chained cryptographic hashing to create an immutable record of the FL training process, enhancing integrity and reproducibility. Each intermediate model state during training is hashed to produce a unique value, forming a chain where each hashed value is dependent on both its current model state and its previous hash. This chain verifies the training process, as even the slightest alteration would cause a hash mismatch when recreating the training process. 

This approach offers a dependable mechanism for validating the FL process, guaranteeing the credibility of its results when data is audited. By ensuring the integrity and authenticity of the process, we build a higher level of trust in the outcomes produced by the system.

\subsubsection{System Design and Implementation}
We design a new FL system architecture that features a data-coupled design to support data provenance and transparency features. This design distinctly separates the data management functionalities from the core FL system, creating a modular structure that enhances flexibility and customization for the various FL processes. In this architecture, local clients can tailor their FL applications by integrating data subsystems based on their specific needs. Not only does this allow for a more personalized application, but it also promotes interoperability, as different subsystems can be easily interchanged or combined. Each client device maintains its database system so that each client can track its history individually, and prevents issues with data privacy such as training model parameters being shared across local devices. 
This separation of concerns between data management and the FL system contributes to the scalability of the system. As the data subsystem is decoupled from the FL system, it can be independently scaled or modified without affecting the overall FL process. This makes the system more adaptable to varying quantities of data and computational resources, thereby enhancing its robustness and versatility in different scenarios.

\section{Preliminaries and Related Work}
\subsection{Federated Learning}
Federated Learning (FL) has surfaced in recent years as a novel approach to collaborative machine learning (ML), providing solutions to several major problems inherent to more traditional ML systems. Prior to the introduction of FL architecture, primitive ML models were typically trained on a central server using data collected from various sources. This approach, while effective, raised significant concerns about data privacy and efficiency in many use cases, as it requires the transfer of highly sensitive data to a centralized location.

FL was proposed as a solution to these challenges. When it was first introduced in 2016, it laid the groundwork for a new paradigm of machine learning and demonstrated its potential for preserving privacy ~\cite{FedAvg01}. In a turn away from traditional architecture, FL would bring the model to the data instead of sending it to a centralized server; this means that the model is trained on each local device or server, using the training data only available to that client. The model update would then be aggregated in the central server to improve the accuracy of the global model. As a result, the approach facilitates the creation of robust, decentralized machine learning models while certifying that data remains on the local device ~\cite{FedAvg01}. The ability to train models locally allows for decreasing communication overhead, increasing regulatory compliance in regard to data management, and uniquely representing the collective knowledge of all clients ~\cite{GBoard}.

Several works have focused on the optimization of FL algorithms for improved performance and efficiency ~\cite{verticalFL2, localupdate}. For instance, the amalgamation of both privacy-preserving and byzantine-resilient algorithms allows for reliable functionality of FL models even in the presence of a node with malicious intent ~\cite{byz01}. The issue of privacy preservation in FL has been another major area of focus. Using techniques or frameworks such as differentially private ADMM algorithms \cite{diffprivateadmm}, BlindFL ~\cite{BlindFL}, and homomorphic encryption ~\cite{he1, he2} have proven effective in enhancing data privacy in FL, however, they are not without their limitations. Attacks such as model inversion, model poisoning, and others have identified core weaknesses in certain approaches and have abused vulnerabilities in FL systems to backdoor training processes and tamper with the data and reliability of a model ~\cite{challenges1, advancesissues}.

When discussing research conducted in machine learning and artificial intelligence, a major consideration is its implication on a model's fairness, accountability, transparency, and explainability (FATE) ~\cite{provenanceoverview}. While work in fairness and accountability of FATE can be universally applied to ML models, there are unique requirements that need to be met in terms of transparency and explainability in systems specific to FL. This is in part due to the fact that FL presents a relatively new approach to collaborative learning, and thus has had less time to truly explore all avenues of research. Existing literature has proposed various methodologies to increase the authenticity of machine and deep learning models (ML/DL), however, there is much work to be done to create key identifiers and metrics specific to building privacy and transparency in FL systems. 

\subsection{Data Privacy in Federated Learning}
By prioritizing data localization over centralization, federated learning offers several mechanisms for enhanced privacy protection. Inherent to FL, raw data is confined to the local device or server, a departure from traditional methods which necessitate data collection and processing on a central server. This localization of data significantly reduces the risk of data breaches and unauthorized access that could transpire during the communication of sensitive data to the central server. However, there are many concerns unique to FL, such as secure communication between edge devices and the central servers, that are not applicable to a typical centralized ML/DL system. Some works, such as ~\cite{secure1} and ~\cite{secure2}, have touched upon such issues, but a comprehensive solution is still lacking. Similar to our introduction of cryptographic chained hashing reminiscent of blockchain technology for FL, ProvChain, a data provenance system for cloud environments, utilizes blockchain-based technology to integrate "tamper-proof provenance, user privacy, and reliability" into the cloud ~\cite{provchain}. This paper helps ground our assertion that our cryptographic hashing methodologies can improve the transparency and verification of models in FL systems.

Federated learning operates by sharing model updates, which are abstracted from the data but do not contain identifiable information. These updates, comprising the weights and parameters of models across many edge devices, are aggregated on a central server to refine the global model. This methodology ensures the confidentiality of the original data, as the updates do not disclose specifics about the datasets used in training.

Further enhancements of the privacy-preserving capabilities of FL include the integration of techniques such as differential privacy and secure multi-party computation. Differential privacy introduces a controlled level of noise to the model updates, thereby preventing the possibility of reverse-engineering the original data ~\cite{diffprivateadmm}. Secure multi-party computation, on the other hand, facilitates computation over multiple user inputs while preserving each user's privacy throughout the process ~\cite{mpc1}. It is also standard for local model updates to be encrypted for added security layers and reduced risk factors. One major development in FL security has been the introduction of homomorphic encryption ~\cite{he1, he2}, a technique which allows computations to be performed on the encrypted data without decrypting it. This allows for the potential to both aggregate local model updates centrally and make predictions without the need to decrypt the data first.

Overall, FL offers a robust framework for privacy-preserving machine learning with many practical applications. Through the localization of data and the sharing of only model updates, FL ensures data privacy while simultaneously enabling the creation of powerful, decentralized machine learning models. The privacy assurances guaranteed by FL systems are particularly relevant in sectors where the security of highly sensitive data is of utmost importance, including healthcare ~\cite{medical1, medical2}, finance, and telecommunications ~\cite{challenges1}. By facilitating the training of machine learning models on private data without necessitating data transmission, FL enables these sectors to reap the benefits of machine learning while adhering to any data privacy regulations ~\cite{datareg1, datareg2, datareg3}.

\subsection{Demanding Model Transparency}
Model transparency and data provenance are indispensable facets of Federated Learning (FL) systems and, more broadly, any machine learning system ~\cite{provenanceoverview}. These elements significantly contribute to the explainability and validity of the models, which are vital for their acceptance and utilization in real-world applications.

\begin{figure*}[t]
    \centering      
    \includegraphics[width=1\textwidth]{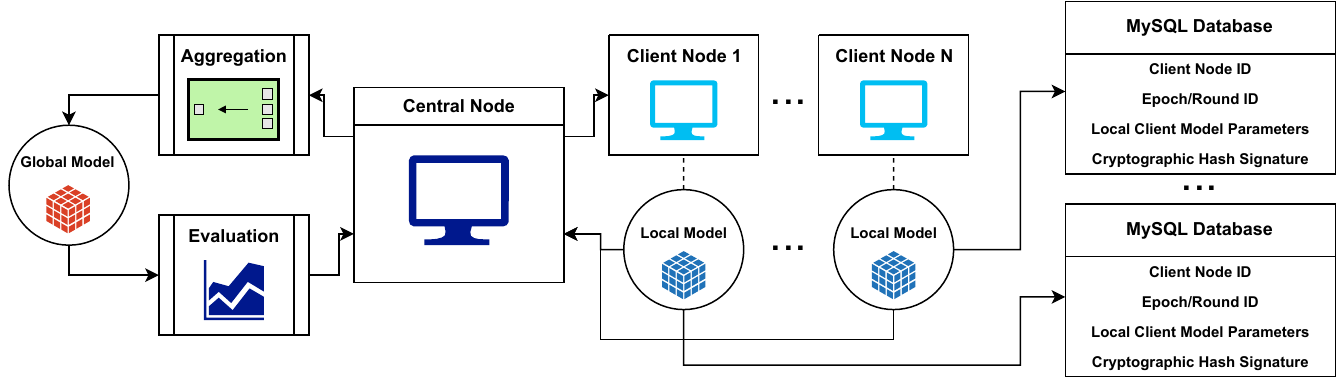} 
    \caption{Proposed Data-Decoupled FL Architecture.}
\end{figure*}

Model transparency, which refers to the comprehensibility and interpretability of the internal workings of a machine learning model, is particularly crucial in the context of FL. Given that FL trains models across multiple devices or servers, transparency guarantees that the global model accurately encapsulates the collective knowledge of all clients. This transparency is of paramount importance in sectors such as healthcare ~\cite{medical1, medical2}, telecommunications, or finance, where model predictions can have great impact and significant consequences.

Data provenance, on the other hand, is the ability to trace and verify the origin of data and its transformations over time, plays a pivotal role in FL. It ensures the reliability and accuracy of the training data the model uses and verifies its appropriate use. Data provenance can also aid in detecting and mitigating issues such as data bias, which can profoundly impact the performance and fairness of the model.

The interplay between model transparency and data provenance also extends to the ethical considerations of machine learning. They are instrumental in ensuring accountability and fairness in machine learning models and can help prevent misuse of data and algorithmic bias. Despite their importance, model transparency and data provenance remain challenging issues in FL due to the distributed nature of FL which can complicate the tracking of each client's model data. Furthermore, machine learning models are often perceived as "black boxes" due to their complex and non-linear nature, which makes the explainability of a model very low.

Addressing these challenges is not only crucial for the advancement of FL but also for the broader field of machine learning. Research in this area can lead to more reliable, fair, and trustworthy machine learning systems. It can also pave the way for wider adoption and acceptance of these systems in sectors where transparency, accountability, and data integrity are key. Therefore, the topic of data provenance in FL is not only significant, but also crucial in order to advance our understanding of these systems, and foster further progression in this research area. These pressing issues underline the importance and benefits of our novel approach, which aims to address these challenges and contribute to the development of more transparent, accountable, and effective federated learning systems.

\section{Proposed Methodology}
In response to the need for data provenance and transparency in federated learning, this paper presents a comprehensive methodology crafted from extensive theoretical exploration and empirical validation. Our approach encompasses a series of innovative features and optimizations to address the challenges faced.

Firstly, we propose a novel architecture that decouples the learning process from the data management. This decoupling is a significant departure from traditional machine learning models, where the learning process is tightly coupled with the data. By separating the two, we aim to improve privacy and scalability, two critical aspects of FL systems. 

Secondly, we introduce strategies to increase transparency and data provenance by systematically storing and managing model parameters as snapshots. These parameters are the heart of the model, determining its behavior and performance. By systematically storing snapshots at the iteration of training, we provide a clear and traceable record of the model's evolution, which significantly enhances model transparency and reproducibility. 

Finally, we employ chained cryptographic hash functions to ensure the integrity and traceability of the learning process. The hash algorithm produces a unique hash value for each input, making it possible to detect any changes or tampering with the data. By storing and chaining the hashed model value at each step, we can verify the model's integrity and increase trust in the federated learning process. 

Together, these strategies target key challenges facing FL and form a robust and all-encompassing approach to improving the efficiency, transparency, integrity, and reproducibility of these systems. Our methodology is not just a theoretical proposition, but a demonstrated practical approach that has been refined through substantial testing and optimization. We explore each data provenance component below.

\subsection{Data-Decoupled FL Architecture}
We propose a data-decoupled Federated Learning (FL) architecture, a novel system approach that distinctly separates the storage of data from the computational aspects. A fundamental shift from traditional FL systems, this architecture, visualized in Figure 2, offers enhanced control over data management and privacy, allowing us to address key challenges associated with model transparency and data provenance, two critical aspects in the realm of FL systems. The motivation for such a system is derived from the necessity to separately leverage the utility of computation and storage. 

The computational side of the model training occurs locally on client devices, taking advantage of the proximity to the training data extracted from the device itself. This allows for efficient training and reduces the latency associated with data transfer to a central server required by non-FL systems. 

On the other hand, the storage aspect is handled by a cloud-based relational database management system (RDBMS). In our implementation, we chose MySQL, a decision driven by several factors that make it particularly suitable for our needs. MySQL offers high performance and reliability, making it an excellent choice for handling storage and auditing the structured data from model snapshots during the training process. It also supports a wide range of data types and provides powerful features for data indexing and querying, enabling us to retrieve specific iteration data quickly and efficiently. MySQL's strong security features, including robust data encryption and access control mechanisms, are essential for maintaining the integrity of our model snapshots and hashed model signatures. 

The database is responsible for storing model snapshots and hashed model signatures at each iteration of the training process. These snapshots provide a comprehensive view of the model's evolution, offering valuable insights into the model's progress and serving as instrumental tools in debugging and optimization efforts. By leveraging the benefits of MySQL, we can efficiently manage and track the model's progress during training without the worry about the security and integrity of our data.

Each model is stored in the database as a collection of the local clients' identifying features, outlined in Figure 3. Specifically, our schema for storing model snapshots is defined as follows:

\begin{enumerate}
  \item \textbf{Client ID:} The Client ID (CID) is a unique identifier for each local client in our database schema. Storing the CID allows us to trace the evolution of a model throughout the training process and track all client devices' contributions. This creates a detailed, comprehensive record of the local clients' learning process which is key for developing a robust data provenance feature that enhances data integrity.
  \item \textbf{Round:} The number of rounds defines the number of times we train the edge devices and aggregate their updates within the central server. Storing this value enables us to create a timeline of a client model's evolution and allows us to snapshot the progression and learned parameters of the model with each iteration of training. This effectively provides us with a timestamp for both auditing and reproducing the training process.
  \item \textbf{Epoch:} The epoch value determines the number of times a local client is trained on its dataset during a single global round. Similar to the round value, this column of our schema provides us with key information to understand the iterations of a client's local model over time, which aids in improving provenance and transparency.
  \item \textbf{Model Parameters:} In our database schema, the model parameters are represented as a dictionary of learned parameters, extracted from a client model. In the context of machine learning, these parameters are the meat of the model which is used to make predictions. Storing model parameters is vital to tracing the evolution of a model over time, providing insights into how the data from each client has influenced the learning process and the global model.
  \item \textbf{Hash Signature:} The hash signature serves as a digital fingerprint of the model at a specific point in time. We use a one-way cryptographic hash function in a chain, similar to a blockchain structure. The storage of hash signatures is crucial for the trustworthiness of the federated learning process as it provides a way to verify the integrity of the model parameters and guarantee that the client has not been tampered with.
\end{enumerate}

\begin{figure}[t]
    \centering      
    \includegraphics[width=0.475\textwidth]{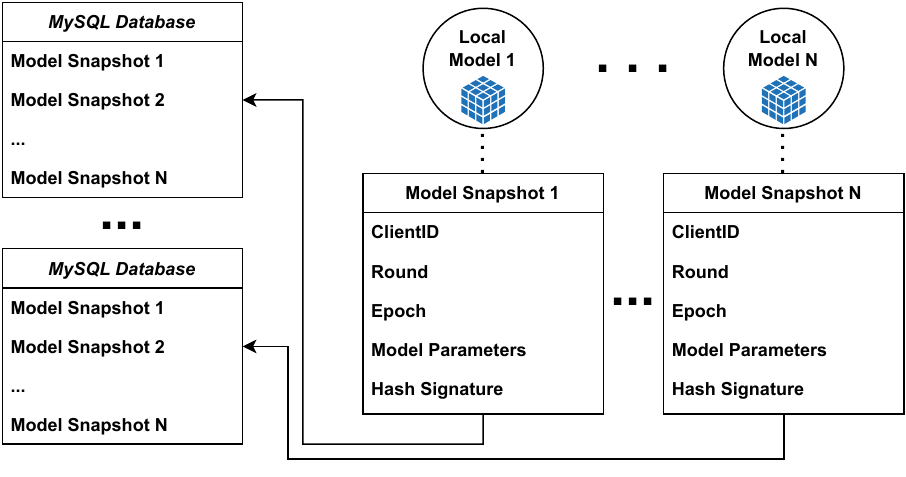}
    \caption{Diagram of model snapshot insertion to a Relational Database.}
\end{figure}

\subsection{FL Transparency with Model Snapshots}
We address data provenance in federated learning systems primarily through the scope of local client model snapshots, which involves the systematic storage of local model parameters at each training round, a practice that offers several benefits and significantly bolsters model transparency. In essence, a model snapshot is the saved state of a model at a specific instance during the training process. It encapsulates the model parameters, which are the internal variables that the model refines through training; these parameters are the core of the model, determining its behavior and performance. In the context of FL, these parameters undergo updates for each participating client node during each round of training on its local data. This local update process is a defining characteristic of FL, facilitating private, decentralized learning.

The necessity to store these local model parameters at each training round is driven by the need for a detailed record of the model’s evolution over time. This record allows us to trace the learning process forward or backward. It provides a granular view of the model's learning process, allowing us to see how the model changes and adapts at each step of the training. Our model snapshot insertion algorithm (Algorithm 1) defines a template training procedure which implements the storage of model snapshots for increased FL transparency. For each intermediate iteration of learning depicted by the nested loops representing global rounds and client epochs, we first train the model, then extract our snapshot. This includes the model's learned parameters, the round ID, epoch ID, and client ID, which we chose in order to capture the model accurately without storing highly sensitive or private data, adhering to the principles of FL. The time complexity of this algorithm is $O(N_{global} \times N_{client})$, as we assume a single database insertion takes O(1) time, as does extracting model information. We do not take into account the time complexity of model training, as we aim to solely encapsulate the time incurred by the database subsystem. 

\begin{algorithm}[t]
\caption{Model Snapshot Insertion}
\begin{flushleft}
\textbf{Input:} $\mathcal{M}:$ Model, $\mathcal{D}:$ Database, $N_{global}:$ \# of Global Rounds, $N_{client}:$ \# of Client Epochs \\
\textbf{Output:} $\mathcal{D}:$ Updated Database
\end{flushleft}
\begin{algorithmic}[1]
    \STATE \textit{\% Global training rounds}
    \FOR{$i = 1$ to $N_{global}$}
        \STATE \textit{\% Client training epochs}
        \FOR{$j = 1$ to $N_{client}$}
            \STATE $\mathcal{M} \leftarrow trainModel(\mathcal{M})$ \textit{\% Train the model}
            \STATE $\mathcal{P} \leftarrow extractParameters(\mathcal{M})$ \textit{\% Extract parameters}
            \STATE $clientID \leftarrow getClientID()$ \textit{\% Get client ID}
            \STATE $roundID \leftarrow getRoundID()$ \textit{\% Get round ID}
            \STATE $epochID \leftarrow getEpochID()$ \textit{\% Get epoch ID}
            \STATE $insertSnapshotToDB(\mathcal{D}, \mathcal{P}, clientID, roundID, epochID)$ \textit{\% Insert model parameters into the database}
        \ENDFOR
        \STATE \textit{\% End of client training epochs}
    \ENDFOR
\end{algorithmic}
\end{algorithm}

As we discussed in Section 3.1, the choice of a relational database for storing model snapshots is motivated by the need for efficient data management and retrieval. Relational databases are designed to handle complex queries and structured data, making them an ideal choice for managing the large number of model snapshots generated during the FL process. This allows us to utilize features such as model rollback, which enables auditors to revert the models to a previous state. Rollback is useful when the model updates result in a performance degradation or unexpected behavior, and the proposed data provenance methodologies provide insights that would support the use of such features. 

Overall, the strategic use of model snapshots in relational databases provides a robust mechanism for tracking the evolution of the model, ensuring data integrity, and enhancing transparency and data provenance in FL systems. This approach, therefore, represents a significant advancement in the field of FL, with potential implications for a wide range of applications. It not only addresses some of the key challenges in FL but also opens up new possibilities for managing and understanding FL models.

\subsection{FL Reproducibility with Hash Signature}
When we discuss fundamental challenges with Federated Learning (FL), a major issue that surfaces is that of trust. In response, we propose an approach that leverages the power of cryptographic hashing functions. Our implementation, which specifically employs the usage of the SHA-256 cryptographic function, serves as a test of integrity and verification of the collaborative learning process. SHA-256 (Secure Hash Algorithm 256-bit). It is used to efficiently generate a unique hash value of fixed size from an input, meaning that no two output hashes will be the same. It is also a one-way hash, meaning it is computationally impossible to recreate the input value given a hashed value, which is crucial, especially to the highly sensitive nature of FL nodes and their local model metadata.

\begin{algorithm}[t]
\caption{Chained Hash Insertion}
\begin{flushleft}
\textbf{Input:} $\mathcal{M}:$ Model, $\mathcal{D}:$ Database, $\mathcal{C}:$ Cryptographic Hash Function, $N_{global}:$ \# of Global Rounds, $N_{client}:$ \# of Client Epochs
\textbf{Output:} $\mathcal{D}: Updated \; Database$
\end{flushleft}
\begin{algorithmic}[1]
    \STATE \textit{\% Global training rounds}
    \FOR{$i = 1$ to $N_{global}$}
        \STATE \textit{\% Client training epochs}
        \FOR{$j = 1$ to $N_{client}$}
            \STATE $\mathcal{M} \leftarrow trainModel(\mathcal{M})$ \textit{\% Train the model}
            \STATE $\mathcal{P} \leftarrow extractParameters(\mathcal{M})$ \textit{\% Extract parameters from the model}
            \STATE $paramString \leftarrow convertToString(\mathcal{P})$ \textit{\% Convert parameters to a string}
            \STATE $prevHash \leftarrow getPreviousHash(\mathcal{D})$ \textit{\% Get the previous hash from the database}
            \IF{$prevHash \neq NULL$}
                \STATE $concatString \leftarrow concatenateStrings(paramString,$ $prevHash)$ \textit{\% Concatenate the parameter string and the previous hash}
            \ELSE
                \STATE $concatString \leftarrow paramString$ \textit{\% First iteration of training}
            \ENDIF
            \STATE $newHash \leftarrow C(concatString)$ \textit{\% Compute the new hash using the cryptographic hash function C}
            \STATE $insertHashIntoDatabase(\mathcal{D}, newHash)$ \textit{\% Store the new hash in the database}
        \ENDFOR
        \STATE \textit{\% End of client training rounds}
    \ENDFOR
\end{algorithmic}
\end{algorithm}

In our chained hash insertion algorithm (Algorithm 2), we showcase a high level overview of the training procedure and process of both calculating and inserting a chained hash value into the model. At each client epoch, or intermediate local round, we first train the model which updates its weights locally. We then extract the newly learned parameters, converting to a string type. We also extract the previous training iteration's hash, as it allows us to chain each hashed model value together. If the previous chained hash does exist, meaning it is not the inaugural round of training, then we concatenate the two string values, and apply a cryptographic hashing algorithm to the result. We finally store our newly hashed model value into our selected database of choice. 

We note that the chained cryptographic hash insertion algorithm runs in $O(N_{global}$ $\times$ $ N_{client})$ time. Similar to Algorithm 1, described in Section 3.2, we do not take into account the complexity of the training algorithm, as we seek to only capture the amount of time incurred by the database subsystem. We also assume that a single database insertion operates in constant time, and that the cryptographic hashing algorithm and model parameter extraction both run in O(1) as well. This indicates that the total time elapsed during the FL process is mostly dependent on the number of training rounds set by the system configuration.

When we chain these hashed model values to one another, we create a structure that is reminiscent of a blockchain, as shown in Figure 4. In a blockchain, each block links the cryptographic hash value of the previous one, creating an immutable, sequential chain. Using this concept as a template, we link our hashed model values, creating a traceable and tamper-proof stamp of the model's evolution. When the training is reproduced during a data audit, the client hash of the reproduction should match that of the original local client exactly, adding an additional layer of security to federated learning. This feature, benchmarked and analyzed extensively in later evaluations, proves to reduce the disk space requirements of model parameters drastically, optimizing the model snapshot insertion time.

\begin{figure}[t]
    \centering      
    \includegraphics[width=0.45\textwidth]{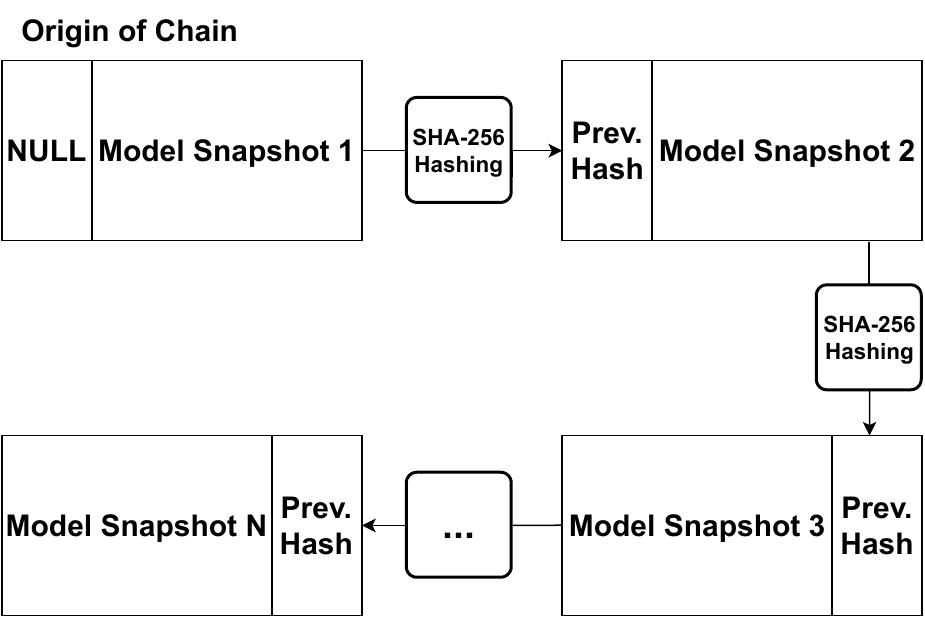} 
    \caption{Hash chain of model parameters}
\end{figure}

\section{Evaluation}
\subsection{Experimental Setup}
We evaluate the performance of our implemented data provenance and transparency features to paint a comprehensive picture of our work and its implications on Federated Learning (FL). Through rigorous benchmarking of our optimizations and techniques, we aim to showcase the benefits of using data-decoupled FL architecture, storing model snapshots, and using cryptographic hashing for blockchain provenance. Our evaluation process encompasses a wide variety of testing environments, including various machine learning model architectures, optimization techniques, and training datasets. All experiments are performed on CloudLab's Utah cluster using the c6525 physical node type, with a 3.00GHz 16-core AMD CPU and 128 GB of memory.

\subsubsection{On Selecting Training Datasets}
To ensure a proper survey of federated learning processes, we selected three distinct image datasets: CIFAR-10, MNIST, and CelebA. These datasets are widely recognized within the machine learning sphere and have been extensively used in research, making them reliable baselines for testing. The diversity of these datasets has allowed us to evaluate our system's performance under a wide range of scenarios. We describe each dataset here: 

\begin{enumerate}
    \item \textbf{CIFAR-10 (Canadian Institute For Advanced Research):} CIFAR-10 is a diverse database consisting of 60,000 32x32 color images ~\cite{cifar10}. These images contain 10 different classifications: airplanes, birds, cars, cats, deer, dogs, frogs, horses, ships, and trucks. The dataset is well known and commonly used for benchmarking as its low-resolution image set allows researchers to quickly test various implementations of algorithms or features.
    \item \textbf{MNIST (Modified National Institute of Standards and Technology):} MNIST is a database of handwritten digits from 0 to 9 that contains 70,000 28x28 grayscale images ~\cite{mnist}. This dataset is widely recognized in image processing and machine learning for its simplicity and allows for moderately fast yet accurate benchmarking of our data provenance features and optimizations similar to CIFAR-10.
    \item \textbf{CelebA (Celebrities Attributes):} CelebA is a large, diverse dataset consisting of over 200,000 178$\times$218 color images ~\cite{celeba}. The images in this dataset are built from 10,177 celebrities with over 40 attribute annotations. Simply due to its size and variety of avenues for testing, CelebA is commonly used for benchmarking and testing various tasks for facial attribute recognition and face detection, among others. We make use of this large-scale dataset in a binary attribute classifier.
\end{enumerate}

\subsubsection{On Selecting Model Architectures}
In order to validate our results further, we opted to use two image classifier models, ResNet-18 and Vision Transformer. Each model provided its own use cases for testing, detailed below:

\begin{enumerate}
    \item \textbf{ResNet-18:} We use Pytorch torchvision's ResNet-18 model, an image classifier boasting over 11 million parameters and a file size of approximately 44 MB. ResNet-18 ~\cite{resnet} provides a popular choice of training model due to its relatively lightweight architecture which still delivers high performance in a variety of tasks. Due to its high rate of usage in the machine learning community, our data provenance work can create an impact that can be directly juxtaposed with other studies, making ResNet-18 a great baseline for comparison. Overall, the ResNet-18 model proves to be a sufficient standard that is not overly complex or overly simple but captures the necessity of optimized data provenance in FL systems.
    \item \textbf{Vision Transformer:} The Vision Transformer (ViT) model is one of the core architectures utilized in computer vision today ~\cite{vit}. We use torchvision's ViT\_B\_16 model which utilizes more than 86 million parameters and 330 MB of space. Adapted from the transformer model, widely used in natural language processing applications and generative pre-trained transformer (GPT) models, ViT models give us the ability to test impact in a larger-scale architecture. Not only are ViT models more accurate in various environments, they are also more computationally efficient than their predecessor, the convolutional neural network, which allows us to compare how the ratio between model size and its training time affects the time overhead of our data provenance features for FL.
\end{enumerate}

\subsubsection{Benchmarking Federated Learning Using Single Node Simulation}
We use Facebook Research's Federated Learning Simulator (FLSim) in order to accurately capture the distributed and collaborative nature of FL systems. FLSim implements a single-node FL system that allows operation without the use of extensive computing resources. We integrate our data-decoupled architecture into this simulation, incorporating the multithreading, model snapshot storage, and blockchain provenance features detailed in our methodology. This allows us to evaluate the impact of these features on the performance of our system in a controlled environment.

We conduct all benchmarking on NSF-backed CloudLab ~\cite{CloudLab01} to further test the scalability and robustness of our system. This cloud computing environment allowed us to evaluate our system's performance on bare metal nodes, providing insight into our work's real-world practicality and scalability. The use of CloudLab provided infrastructure to conduct our full suite of testing, allowing us to assess the performance of our contributions comprehensively and providing consistent access to identical hardware configurations.

In our testing, we collect various metrics according to our database schema, however, we also collect information on the training time, which includes our provenance features. We conducted multiple tests with different configurations to compare the performance of our system under different conditions and optimizations. For instance, we compared the performance of our system with and without the use of model snapshots and cryptographic hash signatures as well as with and without various optimizations, such as multithreading. These comparisons allowed us to isolate the impact of these features on the performance of our system, providing clear evidence, if any, of their benefits.

The visualized data, shown in Figures 5-16, demonstrates the isolated effectiveness of our proposed methodologies for data provenance and model transparency. Our work has shown competitive overheads across the board, and notably, the data-decoupled architecture we proposed made data auditing and benchmarking significantly easier, allowing us to analyze our results without manipulating the FL simulation itself. Not only do the results provide strong evidence for the feasible implementation of a data provenance system for federated learning, they also create avenues for further work in optimization and greater security in FL.

\begin{figure}[t]
    \centering      
    \includegraphics[width=0.45\textwidth]{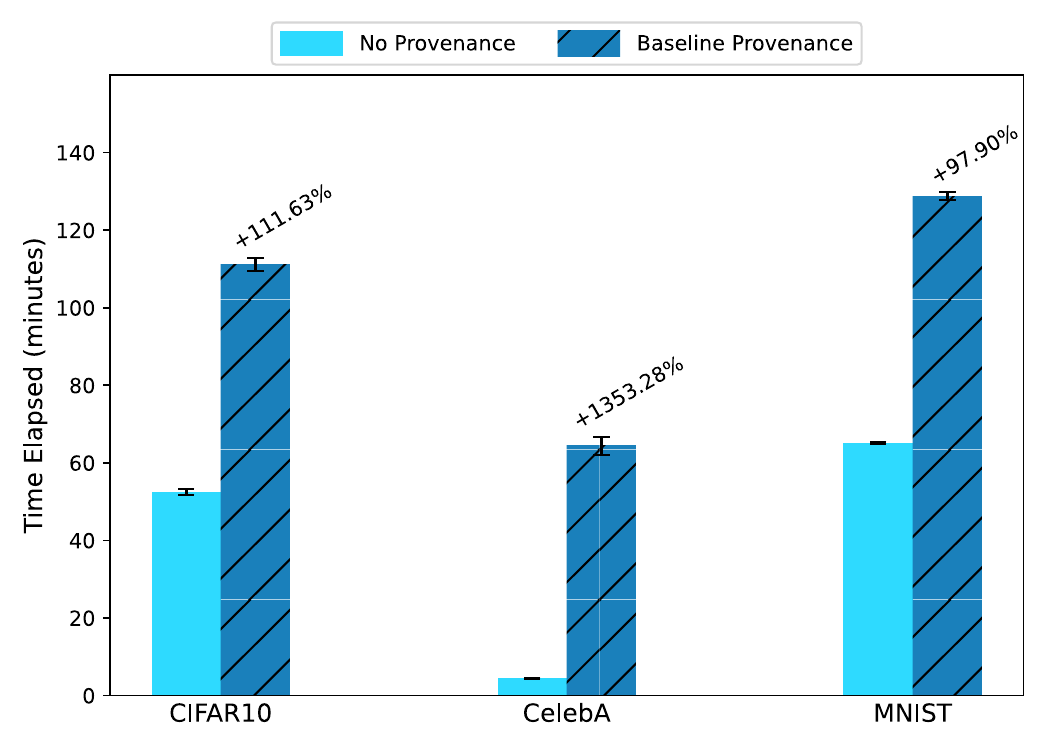} 
    \caption{ResNet-18: Comparison of No Provenance vs Baseline Provenance overheads.}
    \label{fig:resnet_vanilla}
\end{figure}

\begin{figure}[t]
    \centering      
    \includegraphics[width=0.45\textwidth]{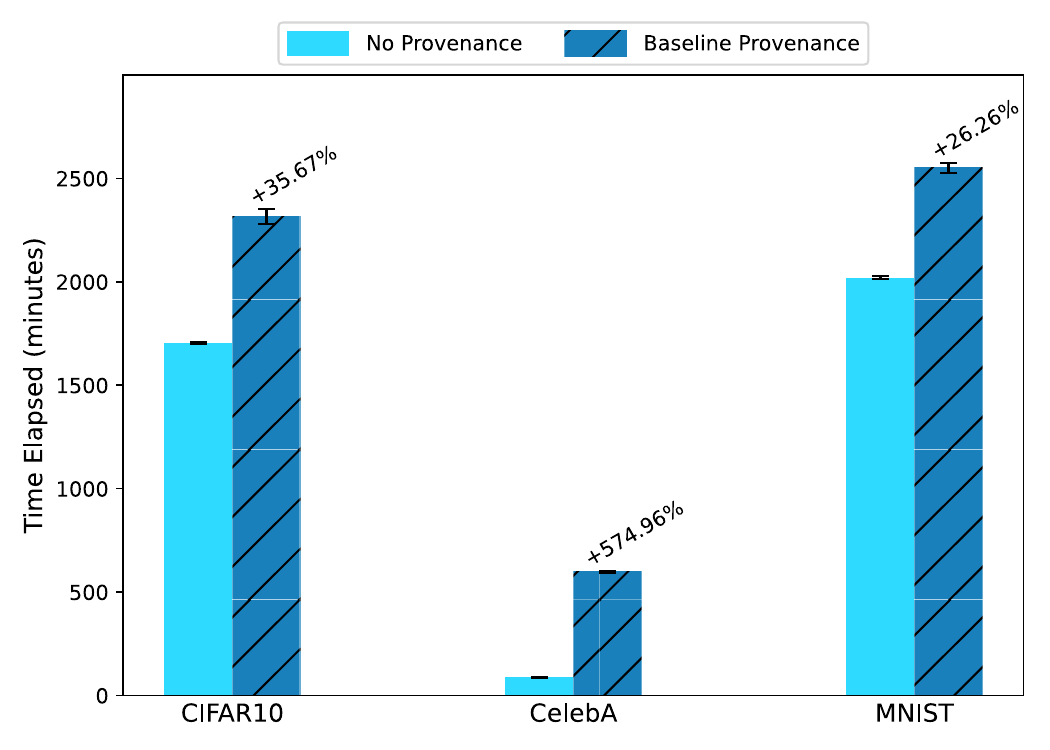}
    \caption{Vision Transformer: Comparison of No Provenance vs Baseline Provenance overheads.}
\end{figure}

\subsection{FL Transparency}
\subsubsection{Baseline Provenance Analysis}
Figure 5 reports the performance overhead incurred by the baseline provenance feature, which records the intermediate model snapshots. The overhead is significant: the CIFAR10 and MNIST datasets finish the training in roughly 2$\times$ period of time and the CelebA dataset incurs 13.5$\times$ overhead. Such significant overhead suggests that a naive database solution for data provenance is infeasible for practical machine learning models like ResNet.

The primary explanation for why the overhead is significant is due to the large number of model parameters and metadata that need to be stored in the MySQL database according to our data schema. We also observe that the ratio of database overhead is higher for CelebA because the training time for CelebA is a small fraction of that for CIFAR10 and MNIST.

\begin{figure}[t]
    \centering      
    \includegraphics[width=0.45\textwidth]{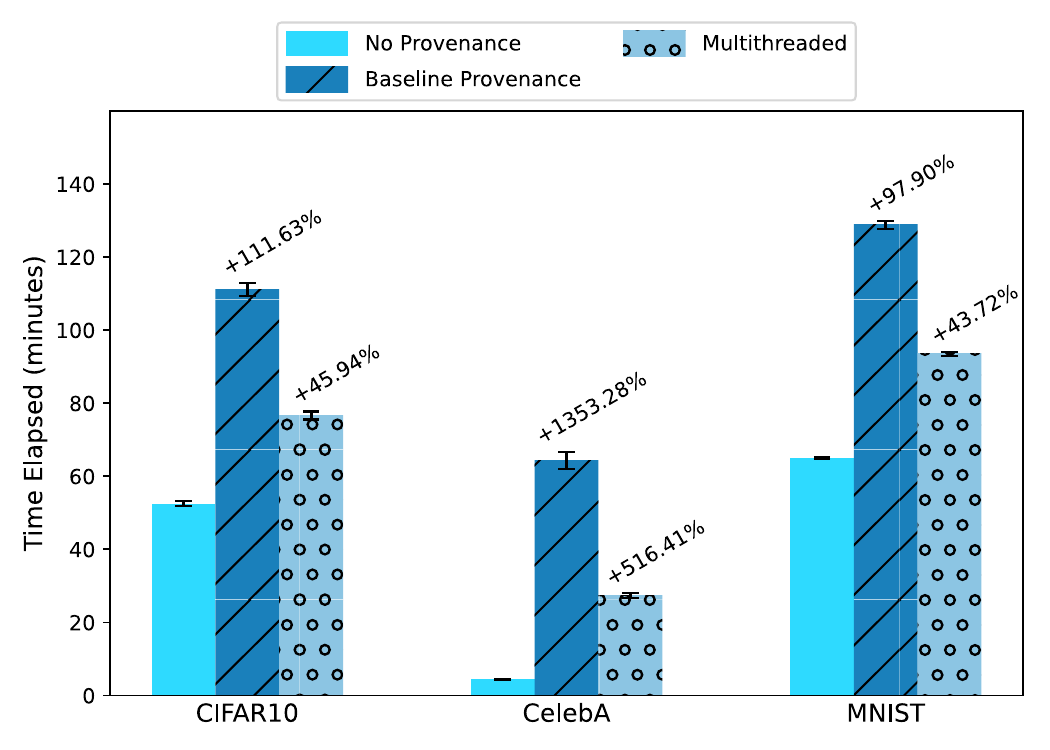}
    \caption{ResNet-18: Comparison of No Provenance vs Baseline Provenance vs Multithreaded overheads.}
\end{figure}

\begin{figure}[t]
    \centering      
    \includegraphics[width=0.45\textwidth]{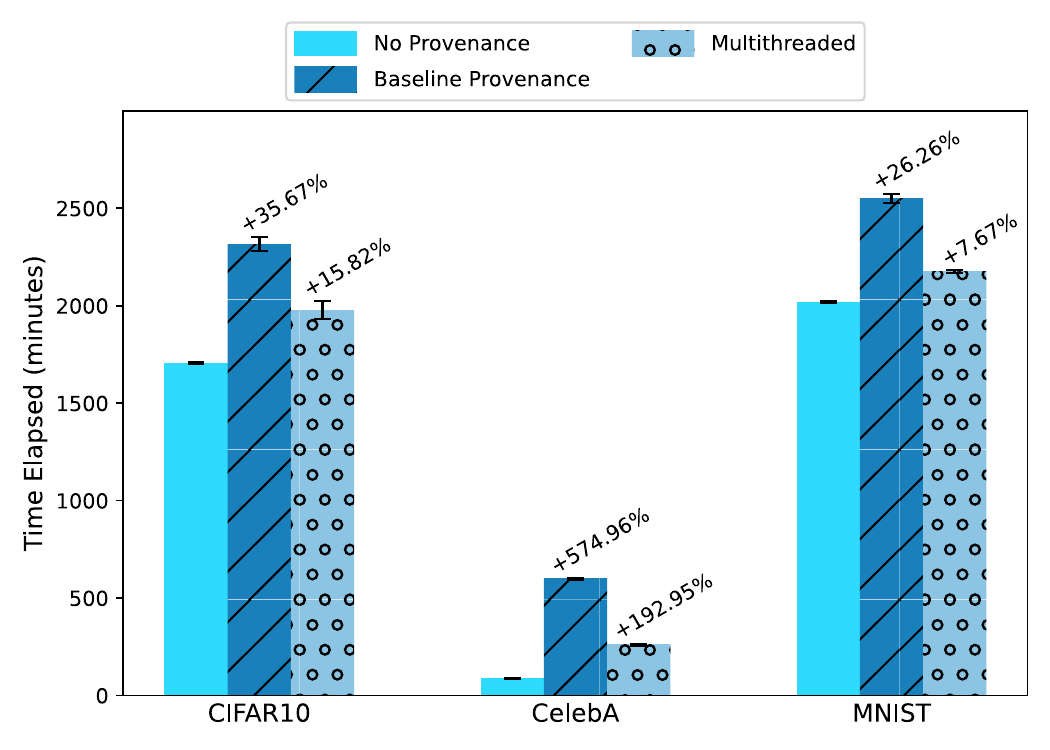}
    \caption{Vision Transformer: Comparison of No Provenance vs Baseline Provenance vs Multithreaded overheads.}
\end{figure}

Figure 6 presents evidence of the high impact of naive baseline data provenance features on model training across the datasets for the Vision Transformer model. The figure depicts a clear comparison that the CIFAR10 and MNIST datasets have a roughly 30\% increase in time. The vanilla configuration incurs a 574.96\% increase in training time compared to a no data provenance build for the CelebA dataset. The overhead is significant because of the SQL injection at each iteration, derived from parameter extraction and storage in an SQL database during each training iteration without any optimization.

Our solution to mitigate these overheads and test comprehensive data provenance and transparency is to benchmark the viability of multithreading, cryptographic hashes, naive snapshot storage, and all combinations of the three. We isolate these tests in order to test individual impact on time overhead, which we use as a measurement for the feasibility of real-world implementation.

\begin{figure}[t]
    \centering
    \includegraphics[width=0.45\textwidth]{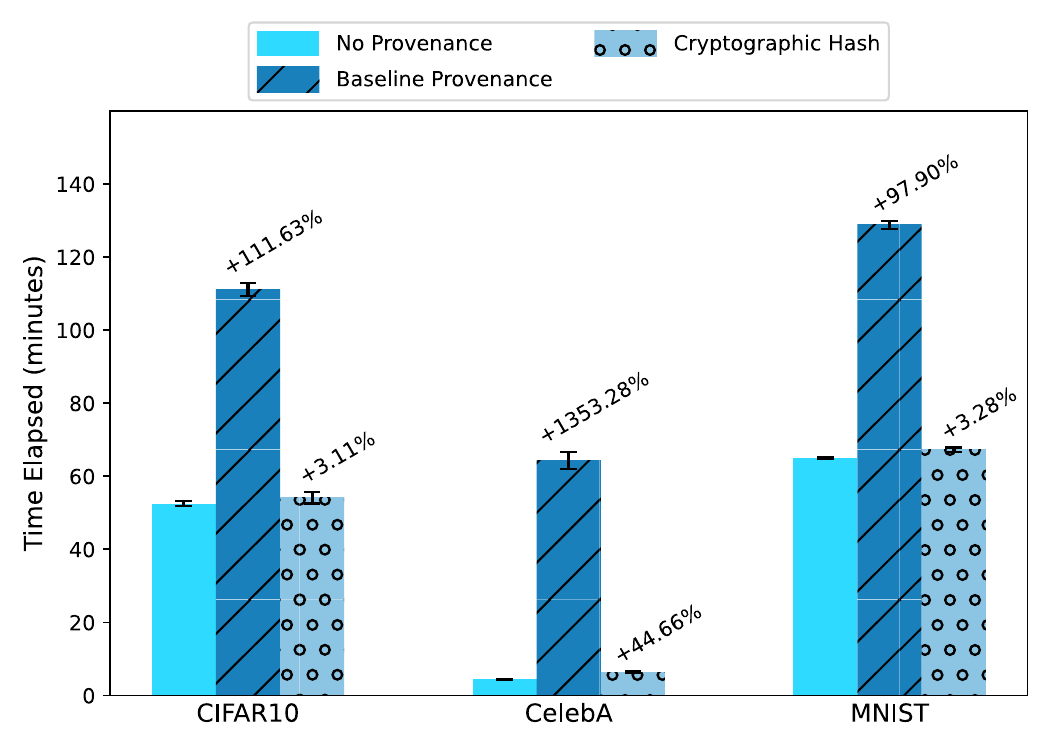}
    \caption{ResNet-18: Comparison of Comparison of No Provenance vs Baseline Provenance vs Cryptographic Hash overheads.}
\end{figure}

\begin{figure}[t]
    \centering
    \includegraphics[width=0.45\textwidth]{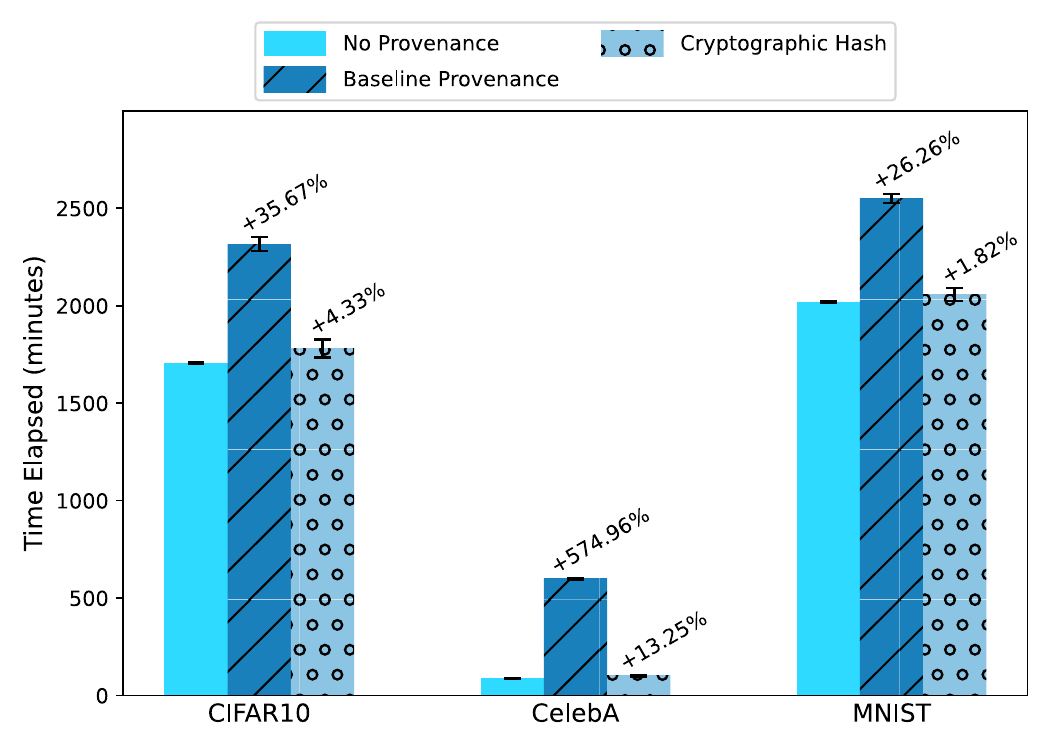}
    \caption{Vision Transformer - Comparison of Comparison of No Provenance vs Baseline Provenance vs Cryptographic Hash overheads.}
\end{figure}

\subsubsection{Multithreaded Provenance Analysis}
Figure 7 gives us evidence to prove that optimizing our baseline data provenance features via multithreading is impactful. The graph demonstrates clearly that multithreaded model training time does not equal the time taken by no provenance training, significantly reducing the overhead time comparatively to the vanilla model by almost 50\%. This could be said that by parallelizing the tasks during the model training and distributing the computational workload of snapshot storage across multiple threads, the processing time can be optimized greatly. For the ResNet-18, a computationally intensive model, this ensures the better utilization of available resources and helps achieve maximum efficiency of the hardware. 

Figure 8 provides a comparative overview of the Vision Transformer model with a baseline provenance system optimized via multithreading. The graph depicts that the datasets CIFAR10 and MNIST have a similar overhead pattern for all three scenarios; the overhead change is roughly 20\% change between the CIFAR10 and the MNIST datasets. The CelebA dataset, however, showcases an overhead difference of 300\%. This is a significant value comparatively considering that CelebA takes a shorter amount of time to train, which increases the overhead impact that storing a model has in this specific scenario. By multithreading, we can effectively store the model snapshot and learn parameters in parallel, which reduces the total amount of time in which the model needs to complete training.

\begin{figure}[t]
    \centering      
    \includegraphics[width=0.45\textwidth]{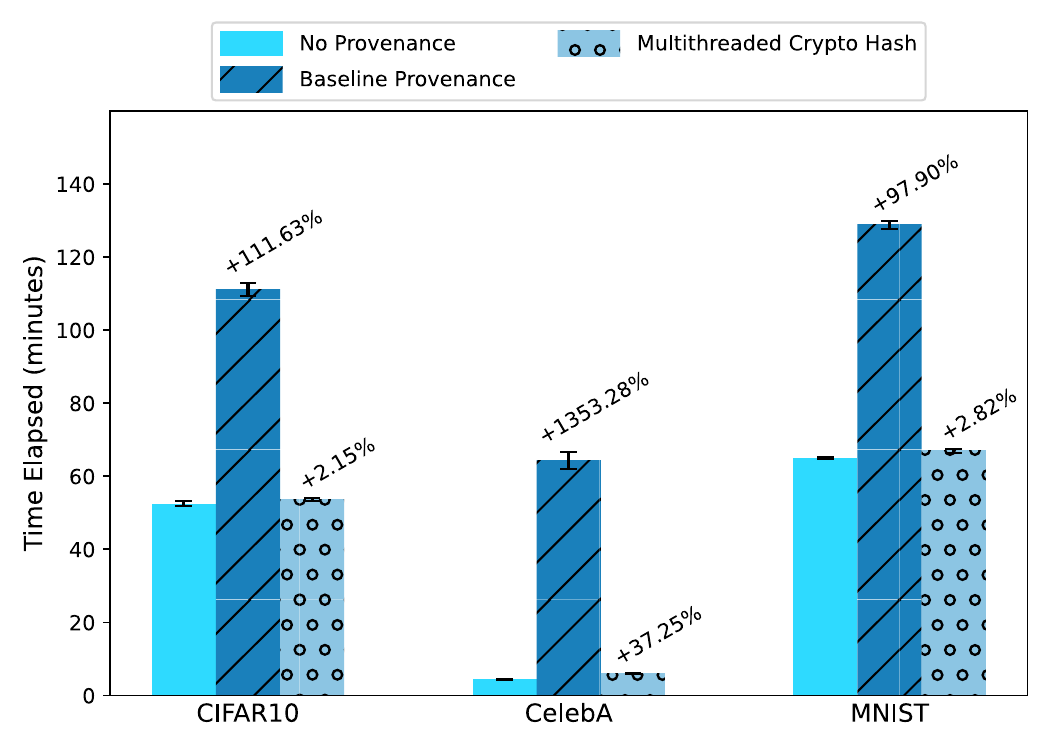}
    \caption{ResNet-18 - Comparison of Comparison of No Provenance vs Baseline Provenance vs Multithreaded Crypto Hash overheads.}
\end{figure}

\begin{figure}[t]
    \centering      
    \includegraphics[width=0.45\textwidth]{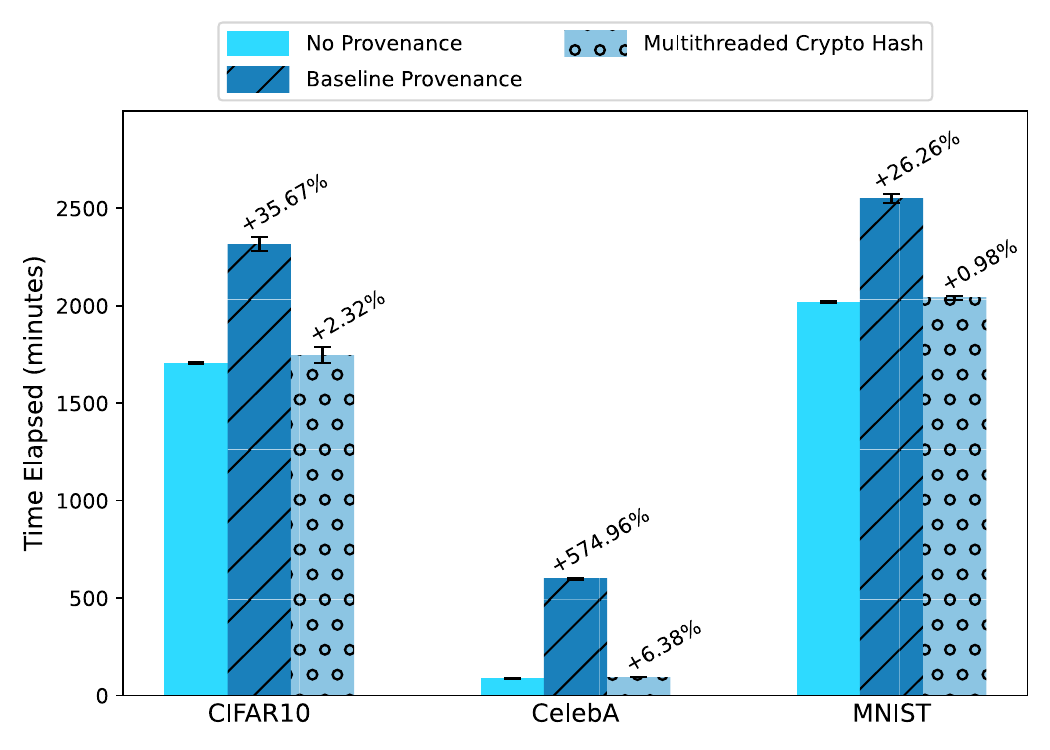}
    \caption{Vision Transformer - Comparison of Comparison of No Provenance vs Baseline Provenance vs Multithreaded Crypto Hash overheads.}
\end{figure}

\subsection{FL Reproducibility}
\subsubsection{Cryptographic Hash Feature Analysis}
Figure 9 compares the performance of a ResNet-18 model across three different provenance methods: no provenance, baseline provenance, and cryptographic hash insertion. While the baseline has an approximately 100\% overhead over the training time with no provenance, the graph indicates that using cryptographic hash insertions decreases the overhead to 3\% for the CIFAR10 and MNIST datasets and around 44\% for the CelebA dataset. This shows a clear drop in the training time due to decrease in the size of the data that is being inserted into the SQL database from creating the chained hashing algorithm. Since the hash function is fast and decreases the size of the data drastically, from 40 MB to 256 bits, the time to inject these values is less than the time to inject the larger, unhashed model.

Figure 10 depicts the impact of our proposed chained cryptographic hashing feature on the Vision Transformer model. The same pattern for the ResNet-18 model can be observed for the Vision Transformer model. The overhead value for the cryptographic hash is notable as it is less than 5\% for the CIFAR10 and the MNIST datasets. The CelebA dataset takes a significant decrease in overhead compared to the ResNet-18 from 46\% overhead to 13\% overhead. The cryptographic hash helps reduce the computational overhead of the training process, and demonstrates its practicality for implementation in a real-world FL system.

\subsubsection{Multithreaded Crypto Hash Feature Analysis}
\begin{figure}[t]
    \centering      
    \includegraphics[width=0.45\textwidth]{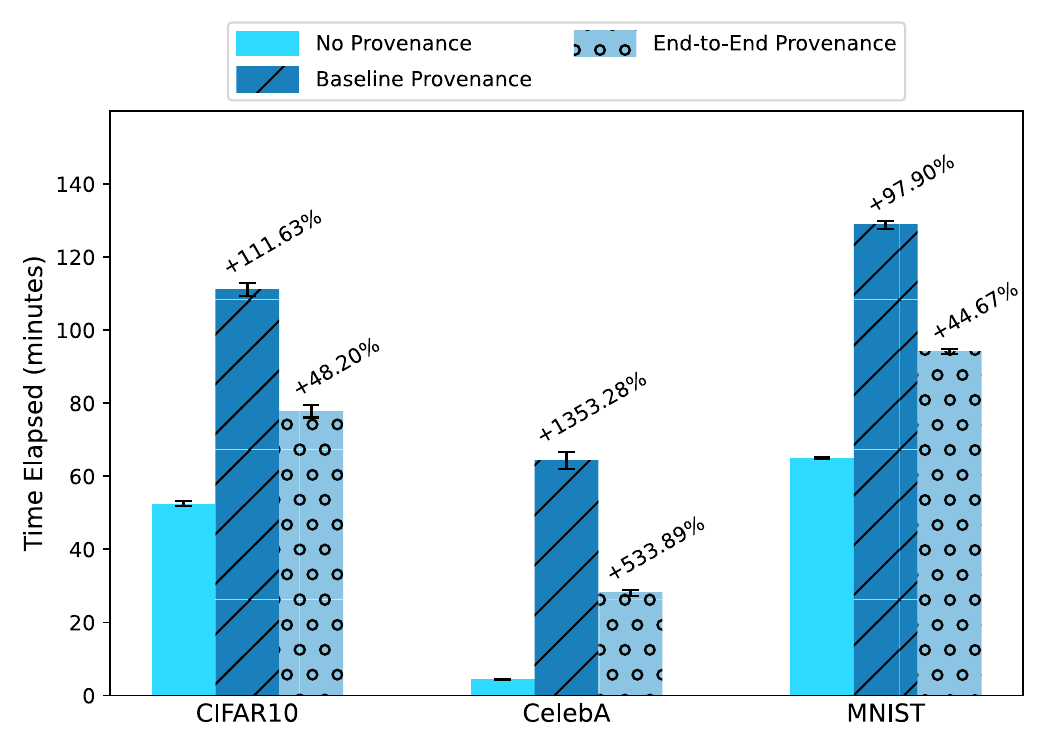}
    \caption{ResNet-18 - Comparison of No Provenance vs Baseline Provenance vs End-to-End Provenance overheads.}
\end{figure}

Figure 11 illustrates the performance comparison for the multithreaded crypto hash provenance feature on the ResNet-18 model. The overhead of the CIFAR10 and the MNIST model is a little over 2\% which indicates that multithreading helped reduce the overhead from an unoptimized cryptographic hash, nearly 3\% for CIFAR10 and 2 \% for MNIST. We similarly see this pattern for the CelebA dataset as well, where by implementing multithreading the overhead was reduced from 44\% to 37\%. This is due to the computational overhead from the cryptographic hash functions being reduced by running the tasks on multiple threads.

Figure 12 evaluates the Vision Transformer for the multithreaded cryptographic hash, confirming the same performance pattern as ResNet-18. By optimizing cryptographic hash insertion via multithreading, the overhead has been reduced by both parallelizing the training and SQL injection as well as reducing the data size of the object we store. We can derive this from the graph that the percentages dropped from 4.33\%, 13.25\%, 1.82\% to 2.32\%, 6.38\%, and 0.98\%, for CIFAR10, CelebA, and MNIST respectively, making it roughly half of the overhead of the plain cryptographic hash overhead.

\subsection{Overall Performance}

\subsubsection{End-to-End Provenance Feature Analysis}
Figure 13 represents the impact of combining multithreaded cryptographic hashing and multithreaded model snapshot insertions for comprehensive data provenance in ResNet-18 model architectures. For CIFAR10, we can see that the time overhead is 48.20\%, roughly equal to the sum of multithreaded (45.94\%) and multithreaded cryptographic hash (2.15\%). This pattern can also be seen for the CelebA and the MNIST datasets. We conclude that the combined overhead is approximately equivalent to the sum of the individual overheads from testing these features. We can also infer that our comprehensive provenance and transparency feature can feasibly reduce additional overhead time by upwards of 50\% for all datasets comparatively to the primitive baseline solution. We demonstrate that our approach helps to enhance the data auditing while maintaining computational efficiency. 

\begin{figure}[t]
    \centering      
    \includegraphics[width=0.45\textwidth]{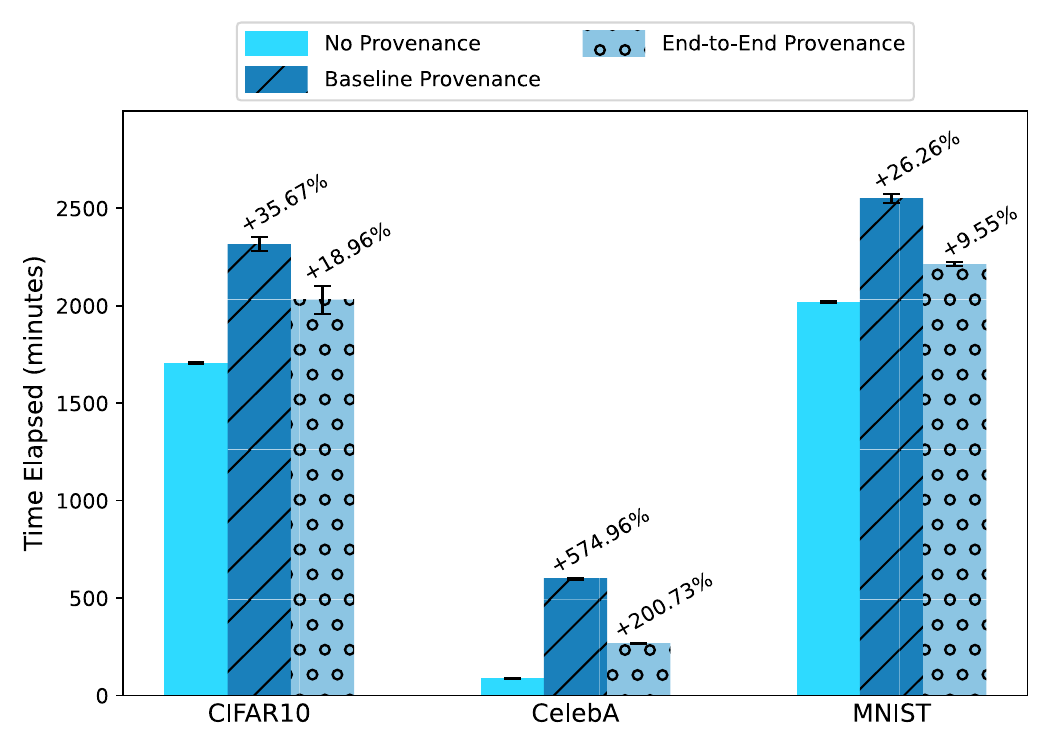}
    \caption{Vision Transformer - Comparison of No Provenance vs Baseline Provenance vs End-to-End Provenance overheads.}
\end{figure}

Figure 14 displays the impact of the combination of multithreaded crypto hash and multithreaded model snapshots insertions on the Vision Transformer model. The pattern again holds from ResNet-18 for the sum of the individual overheads of the multithreaded hash and the multithreaded snapshot insertion. The graph also carries resemblance from Figure 13, where the multithreaded crypt hash and multithreaded snapshot insertion are roughly half that of the vanilla base provenance system, signifying that this system is much more efficient than our baseline. This proves that the patterns followed in Figure 13 and Figure 14 hold across a variety of datasets and model architectures.

\subsubsection{A Comprehensive Analysis of the Data}
Figure 15 combines all the features of the ResNet-18 for comparison. We derive from the graph that the crypto hash feature and the multithreaded crypto hash feature introduces the least additional time overhead, followed by the multithreaded hashing and snapshot insertion feature. As expected, the provenance method has the worst runtime comparatively, demonstrating that our all-encompassing approach to data provenance and model transparency can be implemented without incurring significant time overhead. Figure 16 paints the same image for the Vision Transformer architecture; we can see the same pattern being followed for the graph in terms of how each feature individually impacts the overall performance of the system. 

Our side by side comparison of each isolated contribution shows that our approach offers improved data provenance while simultaneously maintaining computational efficiency, proving it to be a promising choice for privacy-sensitive applications such as federated learning systems. 

\begin{figure}[t]
    \centering      
    \includegraphics[width=0.45\textwidth]{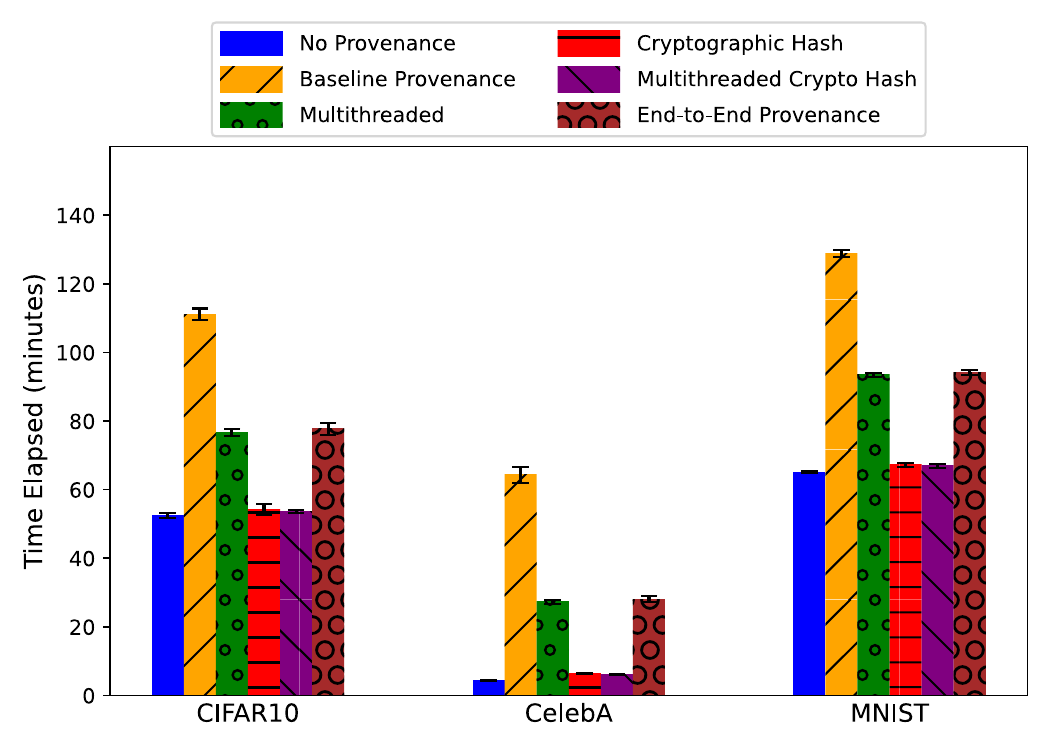} 
    \caption{ResNet-18 - Comparison of all feature overheads.}
\end{figure}

\section{Discussions}
Our evaluations provide strong evidence to support the feasibility and adoption of data-decoupled federated learning with enhanced data provenance and model transparency in many settings. Even after extensive testing of various features and optimizations in our current implementation of the FL architecture, we still find many opportunities for growth and optimization. Due to time limitations, we were unable to explore all potential enhancements and optimizations exhaustively. However, this opens up exciting avenues for further research and development in this field.

\subsection{Impact on Broader Model Architectures} We aim to explore the applicability of our data provenance features on a more inclusive set of models, such as those that use text training datasets for purposes such as sentiment analysis, text processing, and spam filtering, among others. Examining the practicality of incorporating our contributions to these models will provide a comprehensive view of our work's impact on machine learning as a whole, rather than to image classification only. Additionally, text-based models, such as large language models (LLM), have been growing exponentially more popular due to their immense potential for growth and a wide variety of use cases. By extending our data provenance features to a wider variety of machine learning environments, we enable greater versatility and applicability of our work, contributing to the advancement and transparency of ML across all domains.

\subsection{Optimizing Database Systems} In addition to working with other machine learning architectures, we similarly wish to inquire further into optimizing our provenance features via database selection. In our proposed data-decoupled FL system, we utilized MySQL database services to store and manage our data. However, there is much room to consider the possibility of other database systems, including other SQL databases or noSQL databases. Each database offers its own set of unique benefits and characteristics which could potentially offer an improvement in the insertion time of our model metadata. In addition, due to the inherent nature of separation between training and data management, we can easily incorporate various databases and examine their performance in isolation. We hope to further study the impact of each database on the time overhead incurred by our data-decoupled FL architecture and identify which best suits our needs.

\section{Conclusion}
We present a promising approach to improving data provenance and model transparency in federated learning systems, effectively addressing significant gaps in the current landscape of FL. We distinguish our work through the unique application of cryptography for blockchain provenance as well as the optimized storage of model snapshots in our novel data-decoupled architecture. We have established a foundation for improving data provenance and model transparency, and our methodology and approach lays the groundwork for more accountable, secure, and trusted collaborative machine learning. Overall, we assert that our contribution provides improved data provenance and model transparency with a sufficiently low increase in time overhead, but recognize that it opens up new opportunities for further improvement and research.

\begin{figure}[t]
    \centering      
    \includegraphics[width=0.45\textwidth]{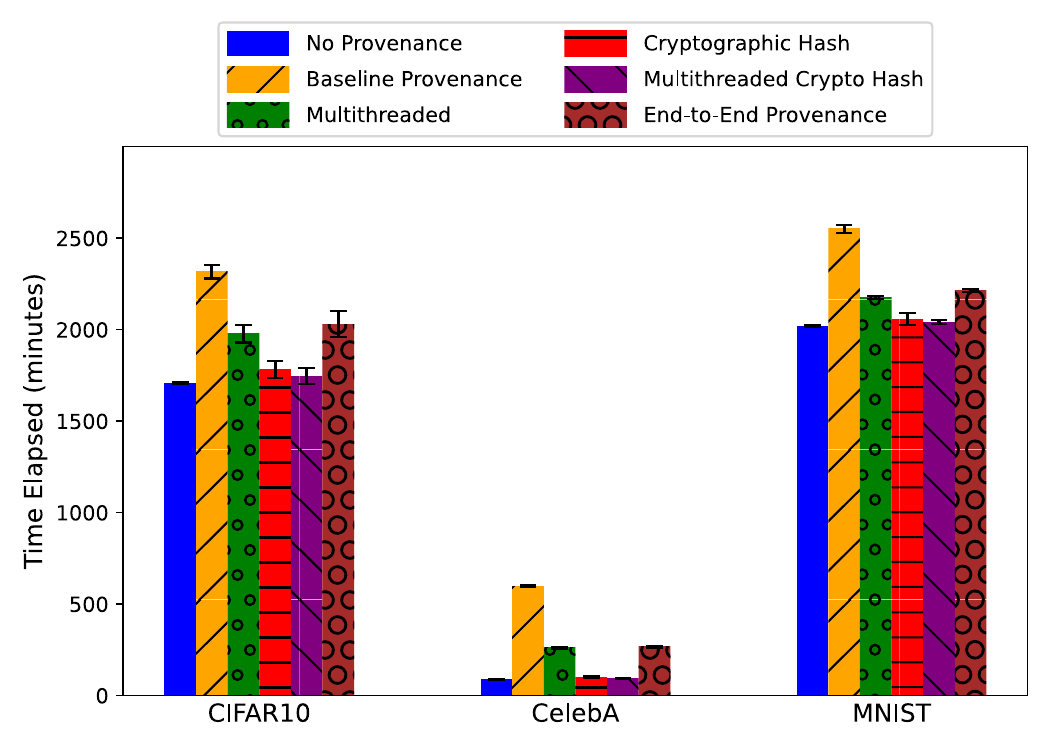}
    \caption{Vision Transformer - Comparison of all feature overheads.}
\end{figure}

\bibliographystyle{ACM-Reference-Format}
\bibliography{main}

\end{document}